\documentclass[%
reprint,
superscriptaddress,
 amsmath,amssymb,
 aps,
prab,
floatfix,
dblfloatfix      
]{revtex4-2}

\usepackage{graphicx}
\usepackage{dcolumn}
\usepackage{bm,amsmath,mathtools}
\usepackage[linkcolor=blue,colorlinks=true,citecolor=red]{hyperref}
\usepackage{float}
\usepackage{silence}
\usepackage{booktabs}
\usepackage{multirow}
\usepackage{tabularx}
\DeclareMathOperator*{\argmax}{arg\,max}
\newcommand{\orcidicon}{\includegraphics[width=8pt]{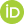}}
\newcommand{\orcidauthor}[1]{%
    \href{https://orcid.org/#1}{\orcidicon}%
}

\begin{document}

\preprint{Submitted to Phys. Rev. Research}
\title{Operational Accelerator Tuning via Model-Coupled Optics and Bayesian Steering}

\author{O. Hassan\orcidauthor{0000-0001-9021-5512}}

\affiliation{TRIUMF, 4004 Wesbrook Mall, Vancouver BC, V6T 2A3, Canada}
\affiliation{Department of Physics and Astronomy, University of Victoria, Victoria BC, V8W 2Y2, Canada}

\author{O. Shelbaya\orcidauthor{0000-0003-1796-3965}}
\email{oshelb@triumf.ca}
\affiliation{TRIUMF, 4004 Wesbrook Mall, Vancouver BC, V6T 2A3, Canada}
\affiliation{Department of Physics and Astronomy, University of Victoria, Victoria BC, V8W 2Y2, Canada}

\author{P.M. Jung\orcidauthor{0000-0002-3409-5365}}
\affiliation{TRIUMF, 4004 Wesbrook Mall, Vancouver BC, V6T 2A3, Canada}
\affiliation{Department of Physics and Astronomy, University of Victoria, Victoria BC, V8W 2Y2, Canada}

\author{O. Kester\orcidauthor{0000-0002-1809-5031}}
\affiliation{TRIUMF, 4004 Wesbrook Mall, Vancouver BC, V6T 2A3, Canada}
\affiliation{Department of Physics and Astronomy, University of Victoria, Victoria BC, V8W 2Y2, Canada}

\author{T. Planche\orcidauthor{0000-0001-9657-4401}}
\affiliation{TRIUMF, 4004 Wesbrook Mall, Vancouver BC, V6T 2A3, Canada}
\affiliation{Department of Physics and Astronomy, University of Victoria, Victoria BC, V8W 2Y2, Canada}

\author{W. Fedorko\orcidauthor{0000-0002-5138-3473}}
\affiliation{TRIUMF, 4004 Wesbrook Mall, Vancouver BC, V6T 2A3, Canada}

\date{\today}
\begin{abstract}
 We present an on-line tuning strategy for the ISAC post-accelerator that pre-sets machine optics with a digital twin and then performs Bayesian optimization for steering under online operation with beam. The model computes end-to-end tunes in seconds and interfaces with the control system under device bounds, slew-rate limits, and loss interlocks. We report three experimental case studies demonstrating that decoupling optics from steering yields faster and more reliable convergence than a fully Bayesian optics-plus-steering baseline under identical conditions. Across these cases, iterations to high transmission tunes are reduced by a factor of 4--6, with final average transmissions in the mid- to high-90\% range. By factorizing optics from steering, the dimensionality of the parameter space is reduced, convergence becomes more predictable, and operational safeguards are easier to enforce. 
\end{abstract}

\maketitle

\section{Introduction}


The design and operation of particle accelerators are shaped by a wide range of scientific, technological, and societal applications. These include nuclear and particle physics studies of nuclear structure and fundamental interactions \cite{Neugart_2017, Svensson2014, doi:10.1142/9789814350525_0003, Jenni2020}, the production of radioisotopes and beams for medical diagnostics and therapy \cite{Robertson2018-mo, LIU2020109154}, materials science investigations using ion implantation and irradiation, accelerator mass spectrometry, and emerging applications in industry and energy research. Across these domains, experiments impose stringent requirements on beam quality, stability, energy control, and species purity. Meeting these demands motivates the development of increasingly sophisticated accelerator modeling, tuning, and control strategies capable of delivering diverse beam species over broad energy ranges with high reliability and reproducibility.

Isotope-Separation On-Line (ISOL) RIB facilities produce and purify rare isotopes in situ, then post-accelerate them with dedicated linear accelerators (linacs) or cyclotrons. Several examples include ISOLDE (CERN) \cite{Catherall_2017}, ISAC (TRIUMF) \cite{Ball_2016}, and SPIRAL 1 (GANIL) \cite{CHAUVEAU202361}. Many heavy-ion linacs employ a low-energy front end consisting of an ion source and a radio-frequency quadrupole (RFQ), followed by structures such as interdigital-H mode (IH) cavities, quarter-wave resonators, or other independently phased accelerating cavities. In a separated-function layout \cite{Laxdal:1997ge}, these cavities and bunchers are individually tunable, offering considerable operational flexibility at the cost of a larger parameter space.

With this system's complexity comes a growing challenge for operators: optimizing beam quality and transmission while taking into account operational overhead. At facilities such as TRIUMF's Isotope Separator and ACcelerator (ISAC), linacs must be re-tuned frequently to accommodate different ion species, each with unique charge-to-mass ratios and energy requirements. Manual tuning is highly variable, time consuming, and places a heavy burden on operators.

The commissioning of the Advanced Rare Isotope Laboratory (ARIEL) \cite{dilling2013ariel, ball2020triumf} will further amplify the complexity of ISAC’s operational scope by tripling its rare-isotope beam availability. ARIEL’s new superconducting electron linac \cite{Laxdal_2016} is designed to produce high-intensity, neutron-rich isotopes for both research and applications, spanning from medical isotope production to generation of intense THz radiation to fundamental science related to dark matter. Together with TRIUMF’s existing 520\,MeV proton cyclotron, the new installation will broaden coverage of the nuclear chart, and once ARIEL is fully online, simultaneous delivery of three RIBs and one stable beam will be possible.

To address this challenge, we present a novel approach to accelerator tuning that decouples the optimization of the beam optical elements from that of the corrective steerers. This enables more intuitive and accessible machine control for operators through rapid, semi-autonomous tuning for beam delivery, although several parameters such as RF phases and ion source settings remain under manual control, these remain the topic of ongoing studies and are outside the scope of this paper. At its core is a digital twin of the linac and beamline that continuously computes beam envelopes and responds to live system parameters. This allows operators to visualize machine states in real time and supports full tune computations from first principles. Combined with a previously reported Bayesian optimizer for beam steering \cite{RSIBOIS}, this method facilitates efficient tune calculation and automated beam threading: the process of guiding the beam through the full accelerator lattice using corrective steerers to avoid losses. This stands in contrast to other methods presently being reported, in which the quadrupoles and steerers are both altered using optimization algorithms \cite{ong:hiat2025-wep12, Awal_2023, bwxw-w9jc, nishi:hiat2025-tup09, GARCES2025165859}. By reducing operational complexity and decision load, our approach improves tuning efficiency and reduces operational overhead for beam delivery. The framework has been successfully implemented at certain sections of ISAC and is now being expanded into several facilities across TRIUMF, including the ARIEL beamline network and the cyclotron injection line. Ultimately, this is an adaptation to the growing demand for more precise accelerator tuning to support TRIUMF’s scientific mission in years to come.

\begin{figure}[!b]
    \centering
    \includegraphics[width=\linewidth]{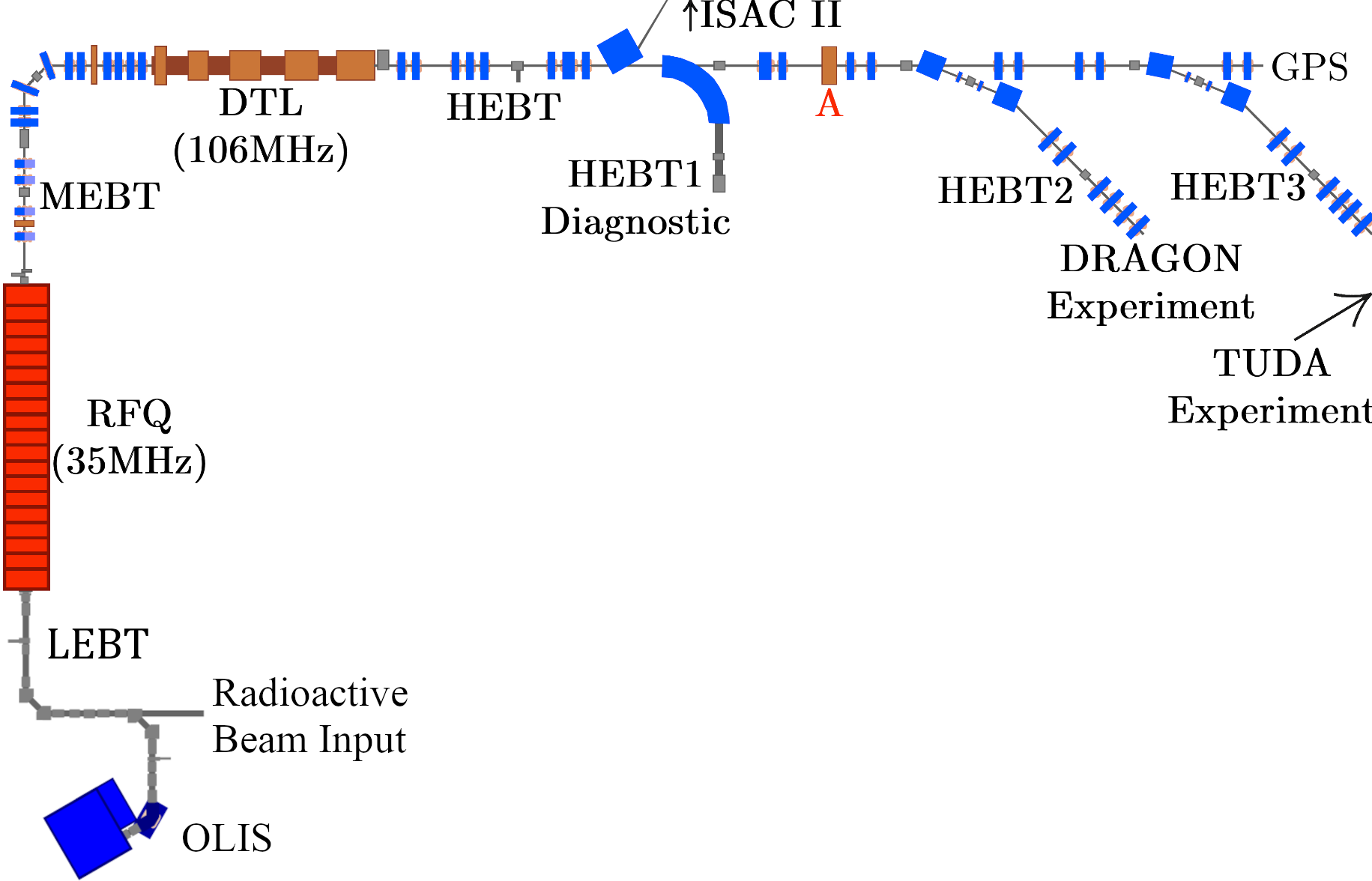}
    \caption{Overview of the ISAC-I RIB postaccelerator, showing major components.}
    \label{fig:isac_facility}
\end{figure}


\subsection*{Relation to prior work}
Prior work \cite{shelbaya2021autofocusing,shelbaya2019fast,shelbaya2024tuning} established fast envelope modeling and sequential optimization within subsections of ISAC (e.g., the DTL) but did not integrate automated beam threading, which was separately reported in \cite{RSIBOIS}. Our contribution is an operational pipeline that (i) computes end-to-end tunes online for the entire ISAC chain with Model Coupled Accelerator Tuning (MCAT) and (ii) applies steering-only Bayesian Optimization for Ion Steering (BOIS) on the live machine. Compared to fully Bayesian optics-plus-steering strategies (surveyed in \cite{roussel} and recent applications \cite{Hassan_2025}), this factorization reduces dimensionality, improves reliability, and aligns with operational safety constraints. We provide multi-species experimental validation, quantitative convergence and transmission benchmarks, and deployment outcomes.

\section{The ISAC-I Facility}

The methodology described in this work builds on extensive operational experience at ISAC, which produces rare isotope beams using proton beam from TRIUMF's 500\,MeV cyclotron to bombard production targets. These are composed of a variety of materials such as UC$_x$, Ta, and NiO for instance \cite{dilling2014isac}. The resulting isotopes are ionized, extracted, separated \cite{bricault2002triumf} and delivered to experiments. Stable beam is provided by the Off-Line Ion Source (OLIS) \cite{Jayamanna2014}. The post-accelerator, shown in Fig.~\ref{fig:isac_facility}, is divided into low, medium, and high energy sections: electrostatic ion optics (before the RFQ) define the low energy section, while magnetic optics (after the RFQ) are used in the medium and high energy sections.

ISAC was commissioned between the late 1990s and mid-2000s, during which time simulation and tuning relied on contemporary beam dynamics codes. This led to a set of modular, piecewise simulation models tailored to specific sections of the accelerator. The low-energy beam transport and matching sections were primarily designed using the envelope-tracking code {\sc transoptr} \cite{baartman2014isac}, while high-energy linac and beamline sections were modeled using ray-tracing tools such as {\sc parmteq} \cite{Crandall:1988uy,TRI-DN-95-04}, {\sc lana} \cite{Junck:1993vc,Ostroumov:1997zz}, and {\sc tracewin} \cite{TRI-DN-99-23}. As a result, tune computations for the entire accelerator have historically relied on multiple independent simulation models. The integration of these into a unified control or modeling framework has recently been completed in the code {\sc transoptr} \cite{shelbaya2021end}.

\begin{figure*}[!t]
    \centering
    \includegraphics[width=\linewidth]{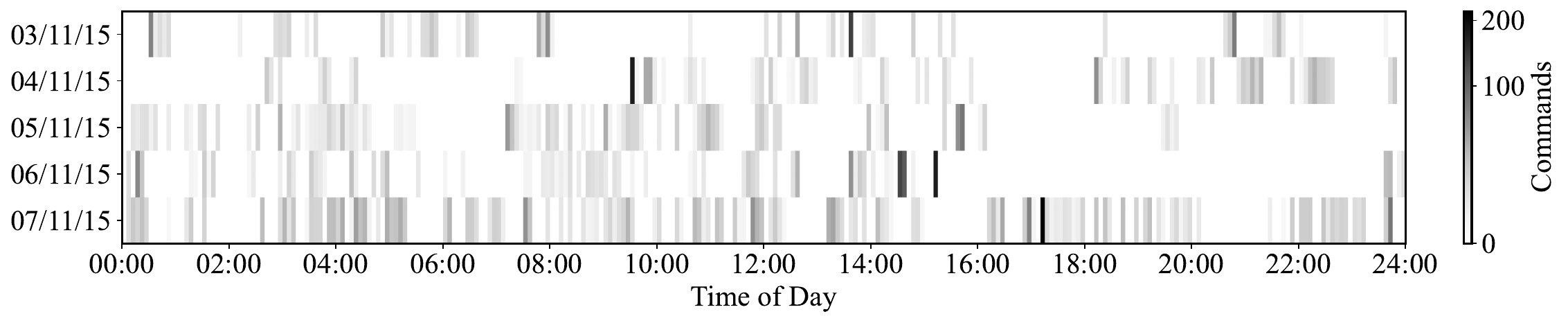}
    \caption{All operator input commands sent into the control system over a 24-hour time period, displayed in 5-minute timebins and shown over 5 consecutive days. In this instance, operators were manually tuning the ISAC-I linac in preparation for scheduled beam delivery. Power law normalization applied for clarity.}
    \label{fig:ECASExample}
\end{figure*}

Historically, beam delivery tunes at ISAC have been stored in two formats: HTML webpages and spreadsheets. Beam tuning initially required operators to manually input device setpoints into the {\sc epics} control system \cite{dalesio1991epics}, consulting various information sources for each setting. This process was both time-consuming and error-prone. Beam from one of ISAC's ion sources would be threaded through the system, from one Faraday cup (FC) to the next, with manual adjustment of transverse steerers and quadrupole lenses to optimize transmission. While this made the tuning process robust and self-optimizing, it posed a challenge on operational resources due to its intensive nature for operators. 

For example, Fig.~\ref{fig:ECASExample} illustrates a typical scenario where ISAC-I is prepared for experimental beam delivery. In the figure, all beam optics and RF control system commands issued by operators are accumulated into 5 minute bins and plotted over five consecutive days; each row corresponds to a day and the horizontal axis shows the time of day. The intensive nature of ISAC post-accelerator tuning can be appreciated by the density of these commands with minimal empty intervals, as operators must continually adjust the machine throughout each day. Automating the tuning of the optics and steerers can therefore offload a substantial fraction of these routine commands and increase operators’ capacity to handle more complex tasks or parallel beam tuning.

ISAC supports a broad program of experiments, namely $\beta$-NMR \cite{Morris2014}, DRAGON \cite{Fallis2014}, GRIFFIN \cite{Svensson2014}, TRINAT \cite{PhysRevLett.120.062502}, and TITAN \cite{Delheij2006}. With the advent of ARIEL and increased user expectations, rapid, accurate, and reliable beam delivery is essential to the facility’s future scientific output. The aforementioned fragmentary nature of simulations complicated whole-machine studies and hindered investigations of emergent phenomena, for example, time-of-flight variations arising from phase instabilities and their propagation through the accelerator lattice, or the propagation of couplings in the beam. We therefore adopt the unified theoretical framework implemented in {\sc transoptr}, as outlined below.

\section{Model-Based Optics Tuning}

\subsection{Linear Optics}

\begin{figure}[!b]
    \centering
    \includegraphics[width=\linewidth]{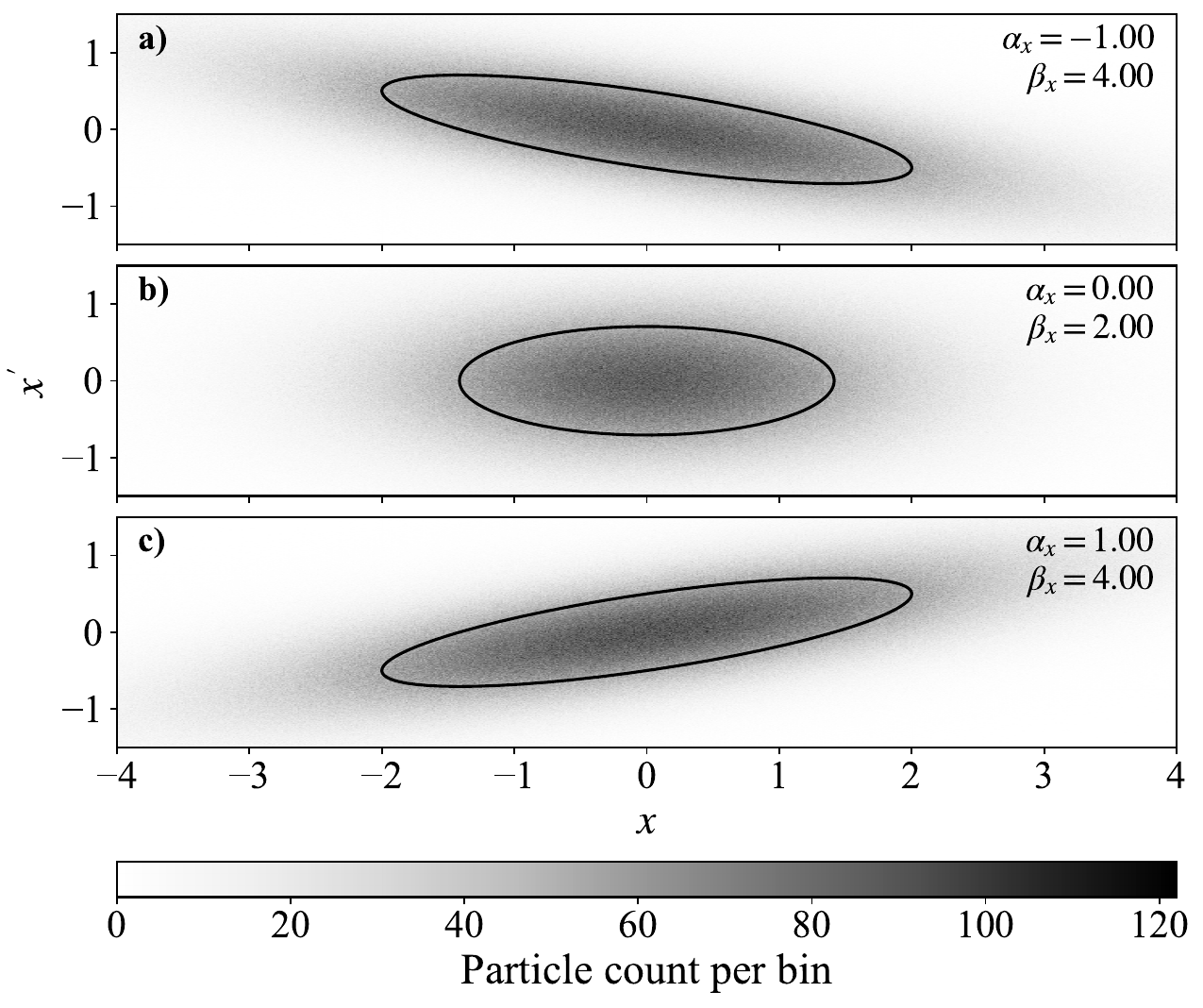}
    \caption{Evolution of a converging beam through a drift. \textbf{a)} Starting distribution where beam is converging $\alpha_x<0$. \textbf{b)} Beam reaches a waist where $\alpha_x=0$. \textbf{c)} Beam starts diverging $\alpha_x>0$.}
    \label{fig:phase_space}
\end{figure}

We adopt the linear-optics formalism familiar from classical optics, but now applied to a six–dimensional phase space. A particle is specified by its state vector $\textbf{X}=(x,x',y,y',z,\delta)$ and the collective beam distribution is described by a \(6\times6\) covariance (or beam) matrix \(\boldsymbol{\sigma}(s)\). The beam matrix contains the second moments of the distribution, and its diagonal elements give the squared beam sizes. The beam envelopes, defined as the RMS sizes, are $\sqrt{\sigma_{11}}$ in $x$ and $\sqrt{\sigma_{33}}$ in $y$. Divergences correspond to the momentum normalized by the total momentum, for example $x' = p_x/P_0$ gives a divergence angle, while $\delta=p_z/P_0$. Instead of time, the longitudinal position \(s\) is our independent coordinate as described in a co-moving Frenet-Serret coordinate system \cite{Wiedemann2015}. The covariance matrix propagates across the beamline through the first-order ordinary differential equations:
\begin{equation}
    \frac{d\boldsymbol{\sigma}}{ds}=\mathbf F(s)\,\boldsymbol{\sigma} + \boldsymbol{\sigma}\,\mathbf F(s)^{\mathsf T},
\end{equation}

where the matrix {\bf F} describes the Hamiltonian dynamics through the local electromagnetic fields \cite{24-38, shelbaya2019fast}. Consequently, any element within the accelerators and beamlines can be modeled in {\sc transoptr} using potentials derived analytically, from field maps, or via finite-element simulations. Computing the infinitesimal transfer matrix at each step gives a prediction of how the beam distribution evolves through the accelerator. Over a small step $ds$, the corresponding transfer matrix is given by $\textbf{R} = \textbf{I} + ds\cdot\textbf{F}$, which will be convenient for formulating constraints in the next sub-section.
The RMS geometrical emittance, $\epsilon_{\mathrm{rms}}=\sqrt{\det\boldsymbol{\sigma}_{2\times2}}$, represents the area of the beam's phase-space distribution \cite{TRI-BN-21-01}. The evolution of the beam matrix is described by a symplectic, first order transfer matrix, if the EM fields are free from higher order aberrations. This means that for optimally efficient beam transport and acceleration through the machine, beams should remain small compared to device aperture sizes and remain close to the central axis through the accelerator. Here, $\boldsymbol{\sigma}_{2\times2}$ is the covariance sub-matrix for a single plane extracted from the full $6\times6$ sigma matrix, either $(x,x')$, $(y,y')$ or $(z,\delta)$, which can be considered independently if there are no couplings beyond the canonically conjugate pairs. The Courant–Snyder (Twiss) parameters \cite{COURANT19581, osti_4430965}, illustrated in Fig.~\ref{fig:phase_space}, are obtained by factoring out the horizontal emittance $\epsilon_{\mathrm{rms}}$ from this $2\times2$ block-matrix; for the ($x,x'$) pair they are defined as:
\begin{equation}
    \beta_x = \frac{\sigma_{11}}{\epsilon_{\mathrm{rms}}} ,\quad
    \gamma_x = \frac{\sigma_{22}}{\epsilon_{\mathrm{rms}}},\quad
    \alpha_x=-\frac{\sigma_{12}}{\epsilon_{\mathrm{rms}}},\quad
\end{equation}
where $\beta_x$ is related to the width of the phase-space distribution, $\gamma_x$ is related to the height, and $\alpha_x$ determines the beam divergence: $\alpha_x<0$ beam is converging, $\alpha_x>0$ beam is diverging, and $\alpha_x=0$ corresponds to a waist. These parameters provide a compact description of the beam for transport and acceleration, and serve as ideal constraints in optics optimizations.

\subsection{\label{sec:implementation}Implementation}
The backbone of our method is a digital twin of the accelerator: a benchmarked model that is continuously synchronized with the real machine’s tune. This enables online simulations and automated optimization of optical elements (quadrupoles, solenoids, and dipoles) and RF parameters. It possesses sufficient predictive power to compute operational tunes during beam delivery. Because the model runs in parallel with the physical machine, any discrepancy between simulated and measured values can be observed by a physicist, signaling miscalibrations in the model or changes in the machine configuration. The resulting feedback loop drives continuous model refinement.

The basis of the digital twin is a unified model of the entire post-accelerator system, which was developed using the beam envelope code {\sc transoptr}. Continuously developed at TRIUMF since 1984 \cite{TRI-BN-16-06}, the code has advanced to handle time-dependent elements whose transfer matrices are not analytically known {\it a priori}. It now supports the inclusion of RFQ \cite{shelbaya2019fast} and multigap accelerators \cite{shelbaya2021autofocusing, shelbaya2024tuning, paulDWA}, allowing for the continuous tracking of the beam and transfer matrices along the structure. {\sc transoptr} allows parameter optimization through user-defined constraints. Input parameters such as quadrupole strengths, drift lengths, and RF amplitudes can be optimized by imposing constraints on the beam matrix, the transfer matrix, or the reference particle’s energy or time. The present version includes two optimizers: Downhill simplex and simulated annealing.

With the Model-Coupled Accelerator Tuning (MCAT) web application \cite{mcat}, {\sc transoptr} is deployed for routine control-room use. To minimize the computational latency of online tuning, we employ sequential optimization; the next subsection details this strategy.

\subsection{Sequential Optimization}

\begin{figure*}[!htpb]
\centering\includegraphics[width=1.0\textwidth]{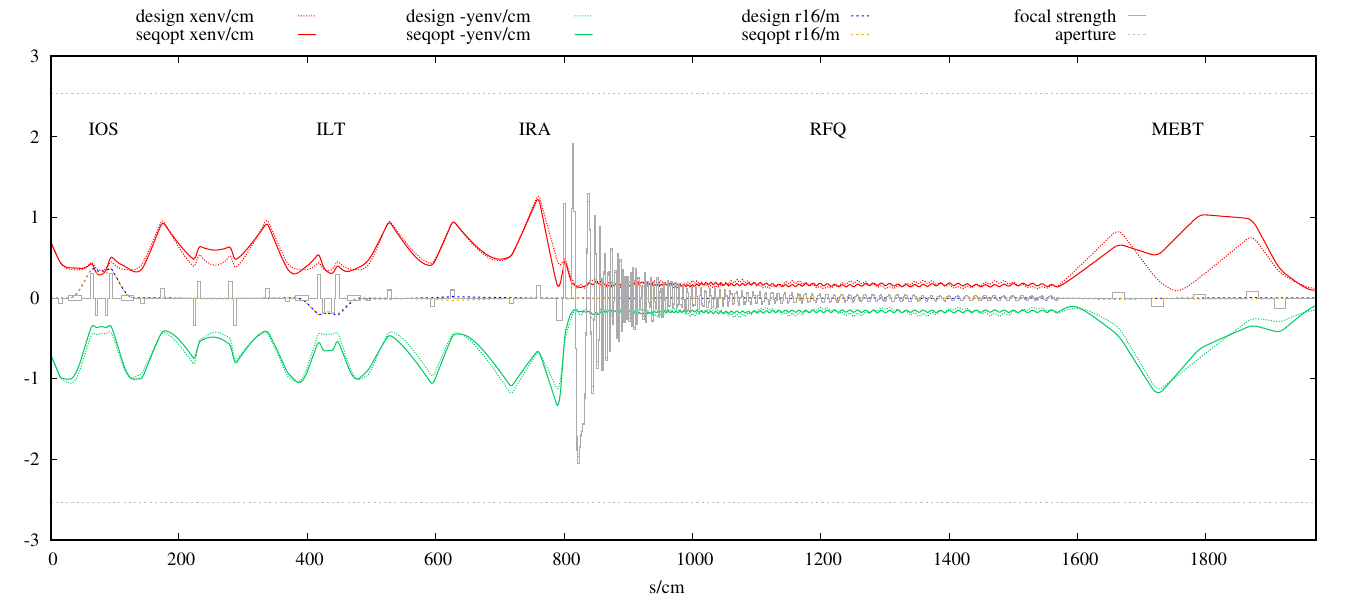}
\caption{\label{fig:tuning_sequences}{\sc transoptr} simulation from the stable ion source line through the RFQ to the MEBT section. The sequential optimization tune is shown alongside the design tune, in solid and dashed lines respectively. The $x$ and $y$ beam envelopes are 2RMS. The focal strengths are scaled up for the quadrupoles and dipoles and scaled down at the RFQ for visual clarity.}
\end{figure*}

Our digital twin uses sequential optimization, where large optimization problems are broken down into smaller and more manageable sequences. Sequential optimization allows for the single execution of the envelope code, which progressively solves each sequence and moves on to the next, while retaining the optimum values found at each step. This way, the lattice can be adjusted online, with the optimization converging quickly. Although for the remainder of this paper, the accelerators and beamlines at TRIUMF-ISAC are shown, the method is general provided the system can be represented by the Courant-Snyder Hamiltonian \cite{COURANT19581}, which in turn means they have known potential functions. The lattice from OLIS to the Medium-Energy Beam Transport (MEBT) was selected to illustrate sequential optimization. This section contains 27 electrostatic and 5 magnetic quadrupoles, since several quadrupoles are wired in parallel, there are a total of 24 independently tunable elements. The optimization was decomposed into nine sequences and achieves a near-identical tune to the design \cite{TRI-BN-12-10}. Table~\ref{tab:tuning_sequences} lists the optimization sequences and constraints, along with the elements varied at each step. While most elements have their initial value kept at 0, some are given an informed starting value based on their typical operating procedure to boost the optimizer. Some constraints are also weighed higher as they are more critical to the optimization. Fig.~\ref{fig:tuning_sequences} compares the design optics with the sequentially optimized tune. Unlike what was reported in \cite{shelbaya2021autofocusing}, which used a {\sc python} wrapper for sequential optimization execution, the method is now implemented natively in {\sc transoptr}’s {\sc fortran} codebase. On an i7-3770S CPU with a single thread and 8\,GB memory, the problem from Table~\ref{tab:tuning_sequences} converges in two to three seconds.



Sequences 1 and 5 tune the achromat sections around each bend. These achromats cancel the dispersion introduced by the dipoles, ensuring that off-momentum particles exit with no position or angular dependence on $\delta$. In practice, the achromatic quadrupoles enforce $\mathbf R_{16}=\mathbf R_{26}=0$ at the exit of a horizontal bend, nullifying transverse-longitudinal couplings which can cause unexpected beam size changes leading to transmission loss. If the dispersion is uncorrected, the focal point of quadrupoles will be smeared in space based on $\delta$, which in turn can create a halo around the beam, compromising the ability to focalize through narrow apertures downstream, for example at the experiment's detector. While it is possible to use sequential optimization to optimize the RFQ, this was not done in this work for simplicity, a static solution was used instead. 


With the development of MCAT and its use of sequential optimization, we can compute end-to-end optics solutions for the entire system in seconds. However, optics alone do not guarantee the beam is transmitted: this requires threading the beam by tuning steerers to compensate for beam alignment errors. To address this, we make use of a machine-learning approach to beam threading, described next (Sec.~\ref{sec:ml_threading}), which computes steerer settings that reliably yield high transmission tunes.

\begin{table}[!b]
  \caption{Beam parameters for sequential optimization.}
  \label{tab:tuning_parameters}
  \centering
  \begin{ruledtabular}
  \begin{tabular}{l c c c}
    Name & Variable & Units & Value \\
    \hline
    Initial energy     & $E_i$          & keV/u & 2.04 \\
    Final energy       & $E_f$          & keV/u & 153 \\
    Mass number        & $A$            & amu   & 20 \\
    Charge state       & $Q$            & --    & 4 \\
    Transverse size    & $x, y$         & cm    & 0.7 \\
    Divergence         & $x', y'$       & mrad  & 20 \\
    Bunch length       & $z$            & cm    & 0.25 \\
    Momentum spread    & $\Delta p/p$   & --    & 0.01 \\
    Correlation $xy$   & $\sigma_{12}$  & --    & -0.92 \\
    Correlation $x'y'$ & $\sigma_{34}$  & --    & 0.92 \\
  \end{tabular}
  \end{ruledtabular}
\end{table}

\begin{table*}[!t]
  \centering
  \caption{Sequential optimization steps from OLIS (IOS) to MEBT.}
  \label{tab:tuning_sequences}
  \footnotesize
  \setlength{\tabcolsep}{8pt}
  \renewcommand{\arraystretch}{1.3}
  \begin{tabularx}{\textwidth}{@{}c l l X@{}}
    \toprule
    \textbf{Seq.} & \textbf{Quadrupole IDs} & \textbf{Description} & \textbf{Optimization Goal} \\
    \midrule
    1 & IOS:Q10--Q13 & First achromat &
      $R_{16}=R_{26}=0$ \\
    2 & IOS:Q9, IOS:Q33 & Fine-tuning achromat &
      $\sigma_{12}=\sigma_{34}=0$ \\
    3 & ILT:Q34--Q41 & Triplet + end quads &
      $\sigma_{12}=\sigma_{34}=0$; $\sigma_{11}=\sigma_{33}<0.5\,\mathrm{cm}$; Match Q34$\to$Q41 \\
    4 & ILT:Q35--Q40 & Triplet only &
      $\sigma_{12}=\sigma_{34}=0$; $\sigma_{11}=\sigma_{33}<0.5\,\mathrm{cm}$ \\
    5 & ILT:Q43--Q46 & Second achromat &
      $R_{16}=R_{26}=0$ \\
    6 & ILT:Q42, ILT:Q47 & Fine-tuning achromat &
      $\sigma_{12}=\sigma_{34}=0$ \\
    7 & ILT:Q47--Q50 & Periodic section &
      Match Twiss parameters for periodicity \\
    8 & IRA:Q1--Q4 & RFQ injection &
      Form waist at RFQ entrance \\
    9 & MEBT:Q1--Q5 & MEBT matching &
      Form waist at MEBT end \\
    \bottomrule
  \end{tabularx}
\end{table*}

\vfill
\begin{figure*}[!htpb]
\hspace*{-1.0cm}\centering\includegraphics[width=1.07\textwidth]{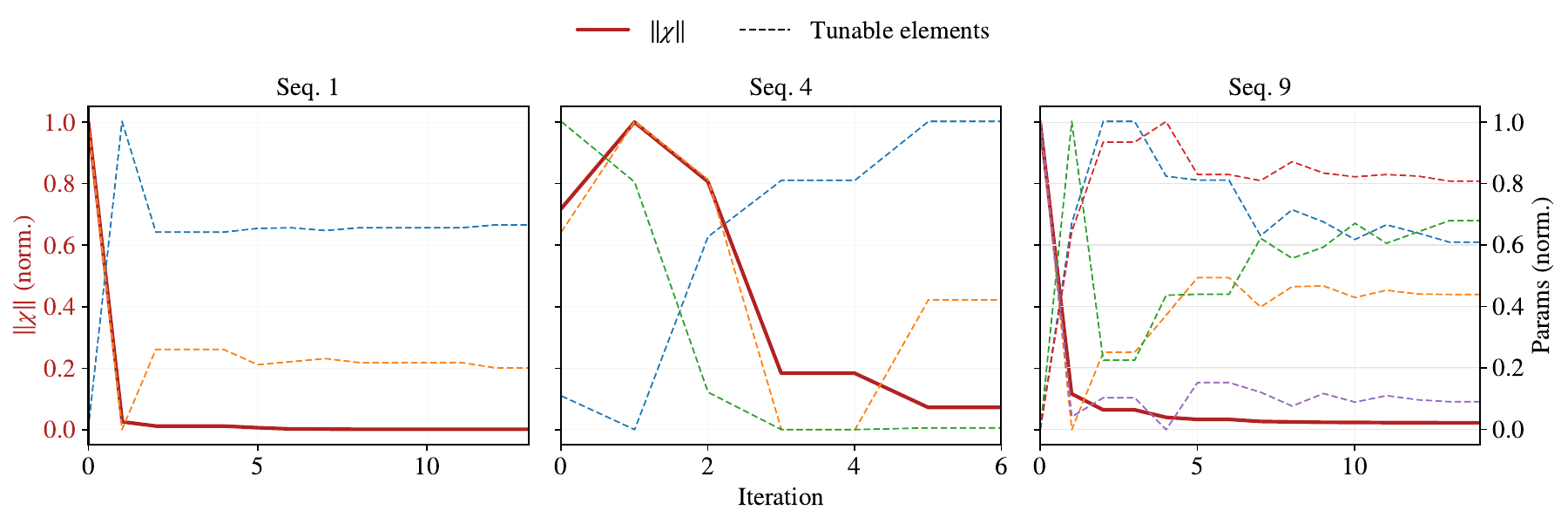}
\caption{\label{fig:tuning_optimization}Optimization history for select sequences. $\|\chi\|$ shown in a red solid line alongside the tunable elements in dotted lines. $\|\chi\|$ was normalized to [0,1]. The tunable elements were normalized using the maximum and minimum of all the parameters.}
\end{figure*}
\clearpage

\section{Steering Corrections}
\label{sec:ml_threading}

The initial centroids of the beam distribution are unknown due to several factors: 1) Lack of diagnostics in the OLIS beam transport section. 2) Mechanical machine misalignments, such as floor-slab settling effects \cite{floor}. 3) Stray fields from both the Multi-Charge Ion Source (MCIS) terminal at OLIS and its steerers, which are affected by steerer lensing \cite{lensing, 24-25}, exhibiting quadrupole-like effects that have been found to be proportional to the sum of the parallel plate voltages \cite{lensingstrong}, which arise due to circular skimmer ground plate apertures together with identically biased opposing steering plates \cite{yuri}. Given the different energies with each delivered beam, the use of theoretical tunes for setting the corrective steerers is not viable. This motivates the use of a machine learning technique for beam threading.

Designed as a complement to MCAT, BOIS provides steering corrections for beam threading. Developed in {\sc BoTorch} \cite{balandat2020botorchframeworkefficientmontecarlo}, BOIS optimizes the corrective steerers while reading back the current from the Faraday cup. The BOIS algorithm has been previously reported in \cite{RSIBOIS} and is used in this work to locally optimize groups of steerers while maximizing beam on a Faraday cup in the line. While this work focuses on one section, BOIS is typically applied in a manner analogous to sequential optimization: the problem can be broken into sequences which allows rapid and efficient local optimizations \cite{Hassan_2025}, from low- to high-energy segments in the machine. 

Bayesian optimization has become a popular tool for many problems in accelerator physics \cite{roussel}. Its main advantage is sample efficiency; it can converge on optimal solutions after only a handful of evaluations, making it an ideal choice for online optimization tasks, such as accelerator tuning and real-time beam delivery. This is a probabilistic technique based on Bayesian inference where a surrogate model, typically a Gaussian process (GP) \cite{rasmussen}, is defined on prior information. The objective is to find the input values that maximize an unknown noisy objective function $\mathbf{x^*} = \argmax_{\mathbf{x} \in \mathcal{X}} f(\mathbf{x})$. 

With a zero-mean prior, a Gaussian process is fully specified by its covariance function, or kernel. The kernel calculates the similarity between pairs of input data points to find the covariance matrix, which determines the posterior distribution. In our implementation, we employ the Matérn kernel \cite{matern1960kernel}, chosen for its flexibility and its ability to directly control function smoothness through the parameter $\nu$. We tested several values of $\nu$, as well as the Radial Basis Function (RBF) kernel for comparison, and found that the most effective configuration used a Matérn kernel with $\nu=5/2$.

\begin{figure}[!t]
\centering\includegraphics[width=\linewidth]{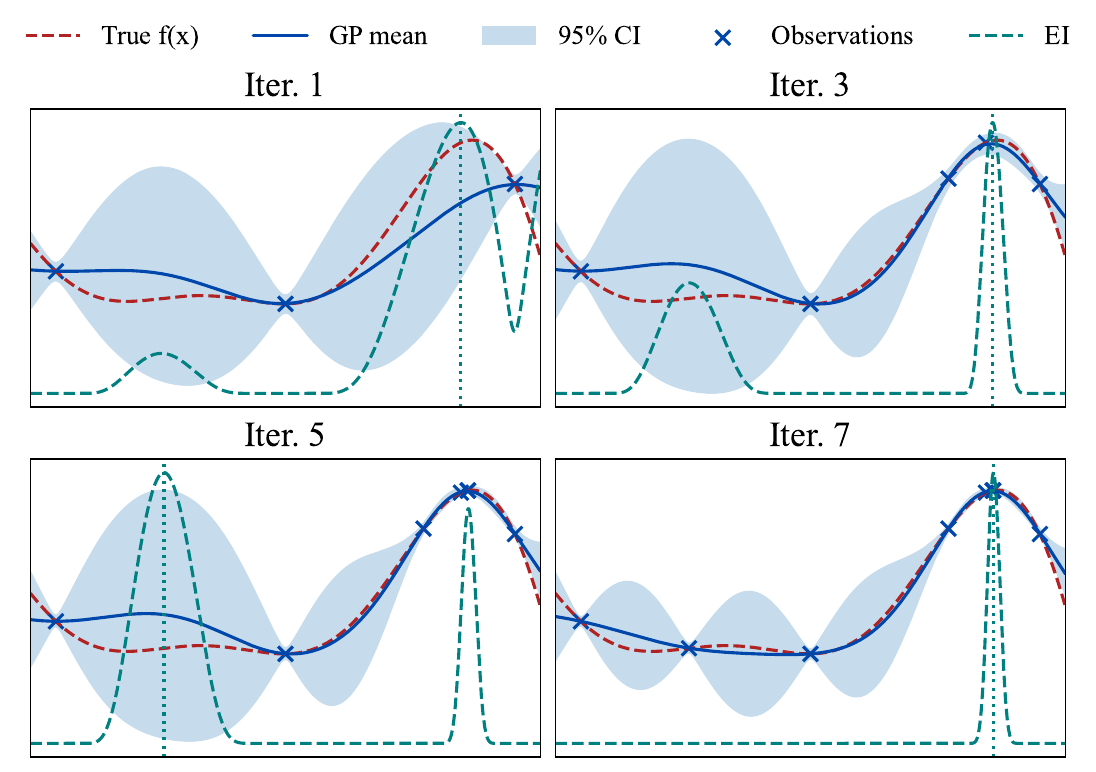}
\caption{\label{fig:gaussian_process}A GP, initialized with 3 random points, is used to model the objective function (red) with sampling guided by the Expected Improvement (EI) acquisition function. The posterior mean is shown in blue, with the shaded band representing the 95\% confidence interval. Acquisition function values are plotted in teal. The horizontal and vertical axes represent inputs (x) vs. function values f(x).    }
\end{figure}

The selection of new evaluation points is determined by the acquisition function (AF), which quantifies the utility of sampling at a given location in the input space, as shown in Fig.~\ref{fig:gaussian_process}. Acquisition functions balance exploration (sampling in regions of high uncertainty) against exploitation (sampling near regions with high predicted objective values). Two of the most commonly used AFs are the Upper Confidence Bound (UCB) and the Expected Improvement (EI) \cite{srivanas2010ucb, ei}, defined as:
\begin{equation}
    \text{UCB}(\textbf{x})=\mu(\textbf{x})+\sqrt{\beta}\sigma{(\textbf{x})},
\end{equation}
\begin{equation}
    \text{EI}(\mathbf{x}) = \mathbb{E}\!\left[ \max\big(f(\mathbf{x}) - f^* - \xi, \, 0\big) \right],
\end{equation}
where $\mu(\textbf{x})$ and $\sigma(\textbf{x})$ are the GP posterior mean and standard deviation, respectively, and $f^*$ is the best observed objective value. UCB favors points with either high posterior mean or uncertainty, whereas EI targets points with the largest expected gain over $f^*$, which combines both the probability and magnitude of improvement. The parameters $\beta$ (for UCB) and $\xi$ (for EI) control the trade-off between exploration and exploitation. Larger $\beta$ or $\xi$ increase the incentive to explore uncertain regions of the input space, while smaller values bias the search near the current optimum. Striking the right balance is important for reliably locating a global maximum. For this work, UCB held at $\beta=3$ was found to be the most reliable AF configuration.

\begin{itemize}
    \item \textbf{Objective \& measurement.} The objective is to maximize the Faraday cup (FC) current. Performance is quantified by the transmission, defined as the FC current normalized to the upstream FC current. Typical measurement noise is on the order of $\sigma \approx 1\text{--}3\%$ of full scale. This noise is treated as a hyperparameter which is learned by the GP.
    \item \textbf{Bounds \& safeguards.} Each element is subject to predefined limits; for example, magnetostatic steerers are restricted to [-100,100]\,A. Power supplies can span these ranges without limitation. A vacuum-loss interlock is continuously monitored throughout the optimization.
    \item \textbf{Zero-Current Crossing.} Magnetic steerers are powered by unipolar supplies coupled with polarity switches, which introduce a finite delay for crossing zero current. For this reason, when BOIS causes a steerer to cross zero, a 5\,s delay is applied to prevent loss of communication with the device \cite{poleswitch}, which is unresponsive during polarity switch.
    \item \textbf{BOIS settings.} The GP surrogate uses a Matérn kernel (\(\nu=5/2\)) with a Gamma probability distribution as the prior \cite{RSIBOIS}. Inputs were normalized and outputs standardized. The random sampling stage was initialized at the midpoints of the parameter bounds \cite{Hassan_2025}. BOIS employed UCB (\(\beta=3\)) or EI (\(\xi=0.1\)) acquisition functions.
\end{itemize}

\section{Results \& Discussion}

We evaluated two tuning strategies on the MEBT to DTL-injection segment, which contains eight quadrupoles and six steerers. In the decoupled method, MCAT’s digital twin with sequential optimization sets the quadrupoles, and BOIS then tunes only the six corrective steerers. In the fully Bayesian method, the optimizer is allowed to vary all fourteen parameters (both quadrupoles and steerers), consistent with practice at other facilities \cite{ong:hiat2025-wep12, Awal_2023, bwxw-w9jc, nishi:hiat2025-tup09, GARCES2025165859}. A stable OLIS beam was delivered in three configurations, and for each method we ran six trials using two acquisition functions: three with EI and three with UCB.

The quadrupole optimization performed with MCAT and BOIS addresses related but distinct problems. MCAT seeks to find the quadrupole settings that best satisfy a set of optics constraints, for example realizing an achromatic bend or a periodic section. By matching these manually defined constraints, MCAT yields beam envelopes that correspond to high transmission tunes through the transport line. In contrast, BOIS directly searches for quadrupole settings that maximize the measured transmission. Although they take different approaches, both methods ultimately aim to maximize beam transmission.

A naive strategy was chosen for the initial sampling stage, where sampling occurs around the midpoints of the parameter bounds for both the steerers and the quadrupoles. In the case of steering, sampling occurs around zero steering which ensures no biasing towards any direction (e.g. no steering beam only to the right). For quadrupoles, the midpoints follow the typical linear optics design. A variable number of sampling points is typically used, ranging from a minimum of $2d+1$ to a maximum of $10d$ \cite{Loeppky01112009}, where $d$ is the number of elements used in the optimization.  However in this paper a fixed number of sampling points based on the minimum $2d+1$ was chosen due to limited beam time.

\subsection{Tuning of the DTL}

\begin{figure}[!t]
    \centering
    \hspace*{-0.5cm}\includegraphics[width=1.1\linewidth]{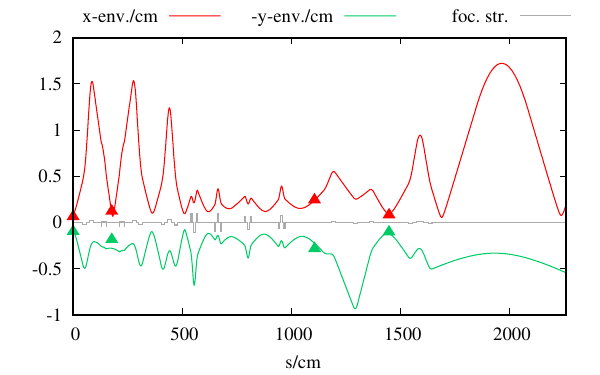}\\
    \caption{Model-computed $x$ and $y$ 2RMS beam envelopes through the ISAC drift-tube linac (solid lines) compared with RMS beam sizes from profile monitors (markers) for a $^{16}$O$^{3+}$ beam. Optics were generated in real time via sequential optimization~\cite{shelbaya2021autofocusing} using an MEBT quadrupole model with fringe-field effects from~\cite{shelbaya2025matchingoptimizationtriumfsrare}. The quantity \emph{foc.\ str.} denotes the quadrupole focal strength. Dataset also used in~\cite{shelbaya2025matchingoptimizationtriumfsrare} for model validation, but shown here to illustrate operational use of the parallel-modeling framework.}
    \label{fig:dtltune}
\end{figure}

Sequential optimization, implemented as reported in \cite{shelbaya2021autofocusing}, is demonstrated on the variable-output-energy DTL. It enables the ISAC-DTL optics to be set instantly for any attainable output energy. This capability removes the need for manual tuning of DTL or HEBT quadrupoles, historically one of the most time-consuming steps in beam delivery. The example shown in Fig.~\ref{fig:dtltune} uses a \(^{16}\mathrm{O}^{3+}\) beam transported through the model-set optics, followed by automated steering corrections, achieving 91\% transmission. The MEBT quadrupole modeling required for accurate DTL injection, incorporating fringe-field effects, is presented in \cite{shelbaya2025matchingoptimizationtriumfsrare}.

\begin{figure}[!t]
    \centering
    \includegraphics[width=\linewidth]{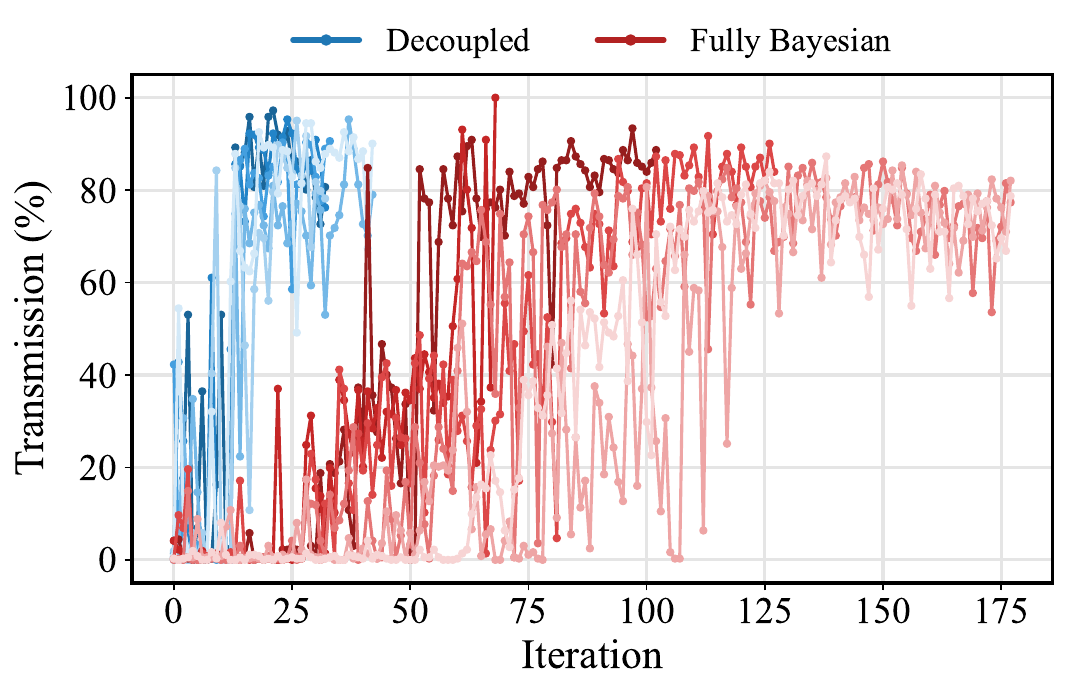}    
    \caption{$^7$Li$^+$ BOIS optimization performance through the DTL for two optimization configurations: steering only (left cluster of points) and steering plus MEBT quadrupoles (right cluster). One step/iteration corresponds to one write-measure cycle for BOIS. Different hues of blue/red correspond to different runs, totalling 6 runs each. Some runs were terminated earlier as they reached a high transmission and were thus considered successful due to limited beam time.}
    \label{fig:BOIS_QUADS_ANDNOT}
\end{figure}

Fig.~\ref{fig:BOIS_QUADS_ANDNOT} compares BOIS optimization through the DTL for two sets of tuning cases. In the first case, all quadrupoles are pre-set using MCAT, and BOIS is applied only to the steerers. Here, convergence to high-transmission tunes is both rapid and reproducible, as indicated by the tight cluster of solutions on the left side of the plot. In the second case, BOIS also varies the MEBT quadrupoles. This step was historically necessary prior to refining the quadrupole model \cite{Hassan_2025, shelbaya2025matchingoptimizationtriumfsrare}, as fringe-field and effective-length errors prevented the decoupled approach. Allowing BOIS to control the MEBT quadrupoles approximately doubles the dimensionality of the search space, leading to slower average convergence and reduced reliability. Because BOIS begins with a random sampling phase, an unfavorable initial sample can leave the algorithm without any high-transmission starting points, forcing it to spend considerably more iterations “hunting” before effective optimization can occur.

\begin{table*}[!htb]
  \caption{\label{tab:opt_comparison}%
    Performance of optimization strategies. Each beam was tested for both fully Bayesian and decoupled cases. Each row represents six separate optimization tests. One step corresponds to one write-measure cycle for BOIS. Conv. time requires transmission to exceed the 1\% threshold to avoid noise biasing the mean, if this threshold is not reached then the conv. time is set to the maximum iteration value.}
  \renewcommand{\arraystretch}{1.15}
  \setlength{\tabcolsep}{6pt}
  \begin{ruledtabular}
    \begin{tabular}{l c c c c c c}
      \textbf{Strategy} & \textbf{Species} & \textbf{Energy (MeV/u)} &
      \multicolumn{2}{c}{\textbf{Iterations to Convergence}} &
      \multicolumn{2}{c}{\textbf{Transmission (\%)}} \\
      & & & \textbf{Mean} & \textbf{Std.\ dev.} & \textbf{Mean} & \textbf{Std.\ dev.} \\[3pt]
      \hline\noalign{\vskip 4pt}

\multirow[c]{3}{*}{\shortstack[l]{Fully Bayesian\\approach}}
  & $^{16}$O$^{3+}$ & 0.436 & 139 & 51 & 80.3 & 25.2 \\
  & $^{4}$He$^{+}$  & 1.60  & 174 & 6 &  4.7 &  8.8 \\
  & $^{7}$Li$^{+}$  & 1.28  & 120 & 34 & 90.7 &  5.6 \\[2pt]
\hline\noalign{\vskip 2pt}

\multirow[c]{3}{*}{\shortstack[l]{Decoupled\\approach}}
  & $^{16}$O$^{3+}$ & 0.436 & 25 &  10 & 94.6 & 3.5 \\
  & $^{4}$He$^{+}$  & 1.60  & 29 &  4 & 98.5 & 0.4 \\
  & $^{7}$Li$^{+}$  & 1.28  & 28 &  6 & 94.7 & 2.2 \\
    \end{tabular}
  \end{ruledtabular}
\end{table*}

These results highlight the advantage of our decoupled strategy: by using MCAT for quadrupoles, with constraints derived from expert knowledge of the facility, we reduce the dimensionality of the problem that BOIS must solve. This leads to faster and more reliable convergence, enabling high performance automated tuning for beam delivery.

\subsection{Decoupling Advantage}

\begin{figure}[!htbp]
    \centering
    \includegraphics[width=\linewidth]{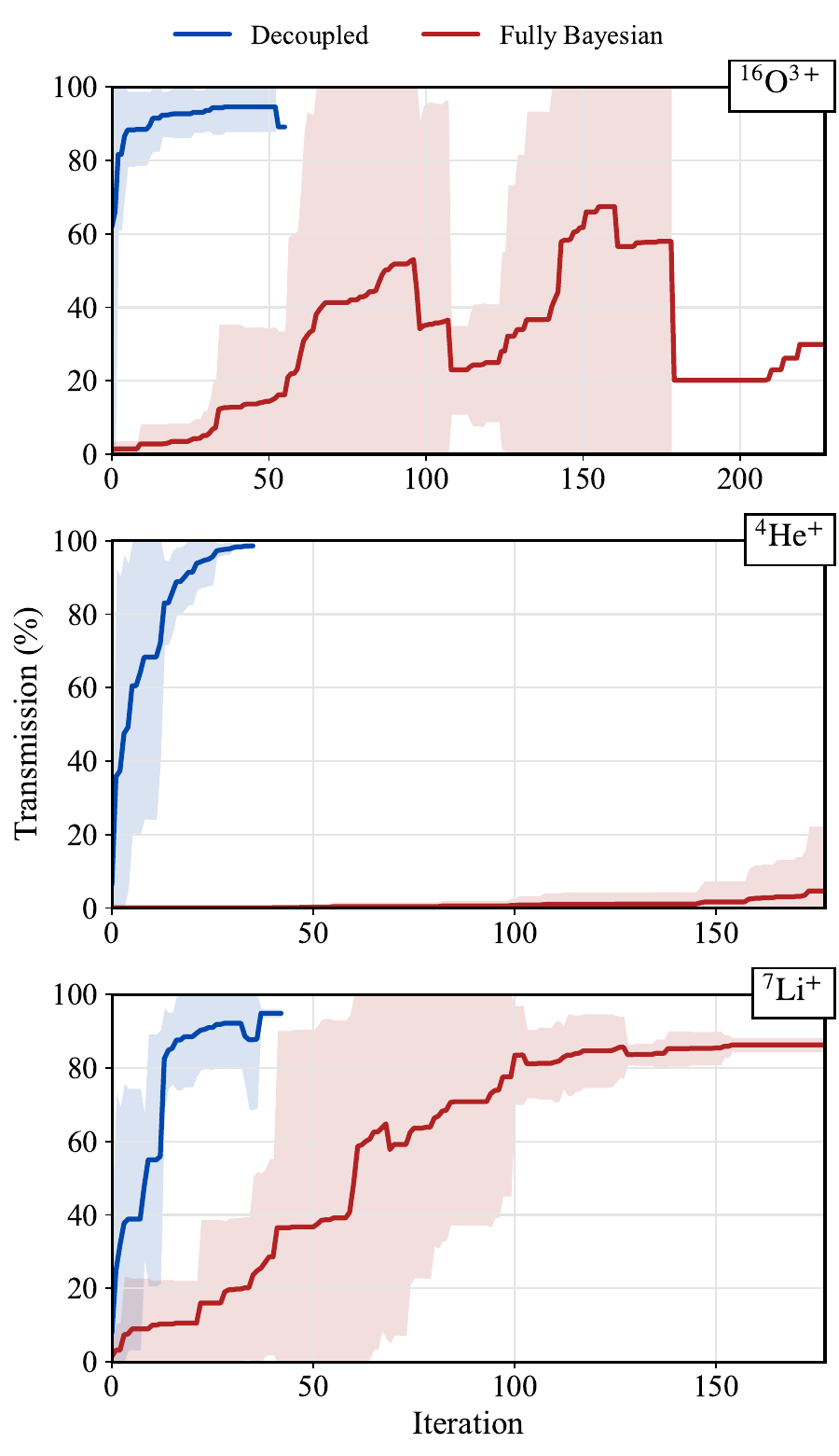}    
    \caption{Optimization performance through the DTL for 3 beam configurations, and two optimization methods: decoupled and fully Bayesian. Each method is averaged over 6 trials. Shaded band shows median and 2$\sigma$ interval across runs.}
    \label{fig:mean_std_bois}
\end{figure}

Table~\ref{tab:opt_comparison} compares the decoupled and fully Bayesian methods. The decoupled method consistently outperforms the fully Bayesian approach: it converges in fewer iterations, achieves higher average transmission, and is more reliable, with reduced variability between runs. Fig.~\ref{fig:mean_std_bois} shows the mean and standard deviation from averaged runs. The $^{16}$O$^{3+}$ test had 3 different convergence regions for the fully Bayesian approach, with the last region only including one trial, as seen in Fig.~\ref{fig:mean_std_bois}. For the $^{7}$Li$^{+}$ test, there was a stripping foil inserted downstream of DTL, leading to an amplified current reading due to the multiple resulting charge states. In this case, transmission was calculated by normalizing to the highest found current across all runs. Together, these results show that the decoupled method is more reliable and efficient for ISAC-linac tuning. 

This improvement arises from several factors. First, as noted above, the decoupled approach significantly reduces the dimensionality of the problem, so the Bayesian optimizer searches a much smaller configuration space for high transmission solutions. Second, continually varying the quadrupoles with BOIS introduces complications due to magnetic hysteresis. In contrast, MCAT mitigates this by using unipolar degaussing \cite{Nasser:2022pud}, which minimizes quadrupole gradient errors. While the same procedure could in principle be incorporated for the fully Bayesian approach, it is impractical due to the time required for each unipolar degaussing cycle ($\sim2$\,min).

Adding quadrupoles raises the dimension from \(d_s\) (steerers) to \(d_s+d_q\). For a fixed initialization size \(k\) and evaluation budget \(T\), the space-filling resolution of the initial sample scales roughly like \(k^{-1/d}\), so increasing \(d\) dilutes coverage per knob; BOIS therefore expends more iterations on exploration (“hunting”) before stabilization, reducing reliability at fixed \(T\).

Another substantial advantage of our method is low diagnostic requirements. No new diagnostics are required beyond those used for standard tuning. The only diagnostics required are a set of beam profile monitors and Faraday cups, which are standard and found in most facilities. The server requirements are also minimal: {\sc transoptr} is CPU-based and Bayesian optimization is more efficient with a CPU as long as the number of training points remains low \cite{roussel}. This sidesteps the need for a GPU which is the most expensive component of machine learning servers.

\subsection*{Statistical summary (small sample)}

All tests done at ISAC occurred during scheduled machine development periods. Beam availability is limited, and the accelerator configuration varies with the preceding experiment. The three case studies were therefore acquired on separate development shifts. Given these constraints we have prioritized comparisons with identical beam compositions but different optimization methodology.

Because we have N=3 such cases (distinct species/energies) with 6 runs per method per case, we have reported per-case effects (convergence time and final transmission) and avoid more involved hypothesis tests. We have compared the proposed method (decoupled) to a baseline (fully Bayesian optics+steering) within each case (species/energy). One \emph{step} is a single write–measure cycle (apply new steerer settings, read the Faraday cup).

Nevertheless, the contrast in Tab.~\ref{tab:opt_comparison} is clear: Across all presented cases, there is a clear advantage to decoupling, with an average of 27 iterations taken to reach well above 90\% transmission through the DTL and into the HEBT line. For the fully Bayesian method, despite significantly more iterations, final transmissions only exceeded 90\% in the $^{7}$Li$^{+}$ case, with both other species underperforming. Within the presented data a clear trend is observed: Decoupling both reduces the configuration space's volume and speeds up convergence to a high transmission final result, owing to the predictive power of the envelope model.

\subsection{Limitations and Safety}
The decoupled strategy assumes correctly calibrated quadrupole strengths. Large calibration errors or strong skew terms can degrade the model optics' predictive power and reduce overall performance. Fast RF phase or energy drifts or high space charge can couple optics and steering; in such cases, local re-optimization may be required. Ion source fluctuations are also currently not monitored by BOIS.

All algorithmic setting changes are constrained by per-device bounds and slew-rate limits enforced at the control system layer. Before each evaluation we verify interlocks and optics device statuses. If a trip occurs, the system reverts to a safe setting by automatically inserting the source Faraday cup. BOIS never writes to radiofrequency cavities, and all optics and steerer commands are constrained by the control system and cannot bypass the safety system. A kill switch allows the user to terminate the algorithm immediately.

\section{Conclusion}

We have demonstrated a novel, hybrid, physics-first approach to accelerator tuning that combines real-time optics modeling with Bayesian optimization for steering corrections. At the core is a continuously synchronized digital twin of the accelerator lattice, implemented in the beam envelope code {\sc transoptr}, which can compute machine tunes in several seconds using sequential optimization. This reduces the tuning problem to a small set of beam steering parameters that cannot be modeled reliably from first principles.

Bayesian optimization is applied only to the steering corrections, addressing the beam threading problem in the presence of alignment errors, fringe fields, and limited diagnostics. Removing the optics from the search space avoids the slow convergence and variability seen in fully Bayesian optimization. Tests on the ISAC post-accelerator show that the decoupled method converges up to about 4-6x fewer steps compared to a fully Bayesian approach, achieves higher average transmission, and is more robust across species and energies.

In operation, the framework turns what was a manual, operator-intensive task into a rapid and repeatable procedure. It requires no additional diagnostics and only modest computing resources, which can be applied during beam delivery without interrupting experiments. The same principles, {\it model where possible and optimize where necessary}, are portable to other accelerators facing high-dimensional tuning problems.

Work is underway to extend automation to other parameters at TRIUMF, in particular RF phase control, which drifts with temperature \cite{shelbaya2018toward} and impacts the longitudinal tune. Integrating these into the same framework will move operations toward a closed-loop, self-optimizing mode suited to high-throughput RIB delivery in the ARIEL era, supported by autonomous tuning methods that the operators can wield at a high level.

\begin{acknowledgments}
We acknowledge the help and support provided by RIB operators during data acquisition. Rick Baartman, Stephanie Rädel, and Spencer Kiy for their advice and useful discussions. We acknowledge the support of the Natural Sciences and Engineering Research Council of Canada (NSERC), [Grant No. SAPPJ-2023-00038]. An LLM was used to refine the writing in some sections. TRIUMF is located on the traditional, ancestral, and unceded territory of the Musqueam People, who for millennia have passed on their culture, history, and traditions from one generation to the next on this site.
\end{acknowledgments}

\noindent\textbf{Data Availability.}
The data that support the findings of this article are available from the authors upon reasonable request.

\bibliographystyle{unsrt}
\bibliography{AllDN,references,ref}

\end{document}